# THE ROLE OF ASTRONOMY IN THE "ANOMALOUS" ORIENTATIONS OF TWO KHMER STATE-TEMPLES

Giulio Magli
School of Civil Architecture, Politecnico di Milano, Italy
Giulio.Magli@polimi.it

*Among the magnificent temple complexes built during the Khmer empire, two single out both for their distance from the Angkor heartland as well as for their "anomalous" - not cardinal – orientation: Koh Ker and Preah Khan of Kompong Svay. Their orientations are shown here to be connected with two relevant astronomical phenomena, namely the zenith passages of the sun and the rising of the Moon at the northern maximal standstill, respectively.*

## 1. Introduction

The Khmer empire flourished between the 11 and the 13 century AD. The heartland of the empire was in the vast Cambodian lowlands, where the kings adopted monumental architecture as a means for the explicit representation of their power, and constructed a series of masterpieces, among which Angkor Wat and Angkor Thom are universally known (Jacques and Lafond 2004).
The first complexes to be constructed are those of Roulos, while later the kings will move to nearby Angkor area, some 15 kilometers to the north. There are, however, two exceptions. One is Koh Ker, located in northern Cambodia some 85 kilometers north-east of Angkor. The other is actually the hugest complex ever built in Cambodia. Usually called Preah Khan of Kompong Svay, to distinguish it from the Angkor temple with the same name, it is located in the Preah Vihear province some 100 Kms east (and we shall return on the meaning of this word "east" in due course) of Angkor.
The function of such architectural ensembles, composed by huge barays (water reservoirs), vast rectangular enclosures enclosing a central temple, palaces, and auxiliary temples and buildings was quite a complex one, as they functioned as royal residence (capital) and main center of cult (state-temple) attesting to the beliefs and religiosity of the king. To complicate matters, the relationship between Buddhism and Hinduism – sometimes exclusive, sometimes sincretist - had a complex interplay with the dedication and the subsequent extensions of these buildings, not to say of later phases of Buddhism/Hinduism explicit conflicts and consequent defacing of temples' images. In any case, construction of the temples was clearly considered as mandatory to attest to the greatness and in some sense to the divinity of the king. The presence of a complex religious symbolism is self-evident, and this is reflected in the planning and in the orientations of these buildings (see e.g. Malville and Gujral 2000, Kak 2001). It is, in particular, well known that almost all the temples, enclosures and barays of Angkor and Roulos are oriented cardinally, with main entrance to the east (except Angkor Vat, which opens to the west). Curiously however, this pattern is *not* respected at the two above mentioned complexes located out of the heartland, so that their orientations can be defined as "anomalous". This fact has been repeatedly noticed in the specialized literature, but never explained.
In this paper we show that these two orientations can be easily interpreted in terms of two very relevant astronomical phenomena: the rising of the sun in the days of the zenith passage for Koh Ker, and the maximal northern standstill of the Moon for Preah Khan. The impressive topographical relationship between Preah Khan and Angkor Vat is also analyzed in this context.

## 2. The orientation of Koh Ker

Koh Ker was the residence of king Jayavarman IV in the mid 10th century AD. The site is characterized by a huge baray and by a 36-meters tall stepped pyramid, which is located in axis with the main temple, the Prasat Thom.
The entire project exhibits a peculiar orientation at azimuth 76° (14° north of east, flat horizon) which is shared also by the short side of the baray. Sometimes topographical reasons - such as the slight south-north slope of the terrain - have been advocated for this orientation (see e.g. Uchida et al. 2014), but it is frankly difficult to believe that the architects of such a huge and complex project might have been influenced by this fact up to rotate the whole design by 14°.
If we search for an astronomical interpretation, an answer is readily found. At the latitude of Koh Ker, azimuth 76° with flat horizon yields a declination of +13° 26'.[1] The latitude of the site is 13° 44' so the complex main axis is quite precisely oriented to the rising sun on the days of the zenith passages, which of course occur when the sun has a declination equal to the latitude of the place (April 27 and August 16).

## 3. Astronomy in the project of Preah Khan of Kompong Svay

Preah Khan of Kompong Svay is composed by a series of four "concentric" rectangular enclosures which contain auxiliary temples and buildings and a central sanctuary. The exterior perimeter of about 5 km per side makes it the hugest Khmer enclosure ever built (Mauger 1939). Of course the complex has its own baray, a rectangular reservoir built along the main axis of the complex. The site was connected to Angkor by a "royal" road very rich in stone structures, such as bridges and "rest house" temples. The building chronology is difficult to establish, since only one dated inscription (1010 AD) has been recovered, in one of the temples of the complex, Prasat Kat Kdei. Accordingly, the site might have been founded in the 11th century by king Suryavarman I. However, important architectural details – and in particular, the central towers – rather point to the first half of the 12th, during which Angkor Wat was constructed. Curiously, yet other details recall the late 12th to early 13th century, pointing to king Jayavarman VII, the builder of the Bayon, the famous face-towers edifice at Angkor Thom. The religious dedication of the complex is equally difficult to individuate, due to the interplay between Buddhist and Hindu elements.
The causes leading to the construction of such a majestic architecture in such a remote place are still subject to debates. There is no doubt that industrial activities related to iron melting were carried out in this area and also inside the enclosure, but of course the presence of such an impressive, symbolic monument remains difficult to explain, and a complete re-analysis of the archaeological setting together with a new mapping of the area with modern techniques is currently giving new insights into this problem (Hendrickson & Evans 2015). As far as we are concerned here, there is an aspect which has been noticed by all authors but never explained satisfactorily, namely – again – the anomalous orientation with respect to the Angkor monuments, in spite of the close architectural analogies mentioned above. The complex is indeed clearly rotated to the north of east.
The orientation is given by Mauger as 27° 24' north of east, but repeated measures on satellite images and with the help of Google Earth rather point to a greater value, close to 30° north of east (azimuth 60°). It is this value which will be used here; the horizon is flat or nearly flat.
The hypothesis which has been analyzed by some authors (see e.g. Paris 1941[2]) is that the complex might have been orientated to the rising sun at the summer solstice. However the azimuth of the midsummer sun (with a flat horizon) in this region is 24° 30' and therefore definitively far from the observed one. An error of several degrees is out of question, considering the skills and technique of the Khmer architects and the fact that, since the error is to the north, the sun does *never* rise in alignment with the temple in the course of the whole year, and it is of course almost impossible to think that such an error might have been generated by solar observations. So, definitively, the temple was not oriented to the rising sun. Of course this does not mean that it was deliberately oriented to another astronomical phenomenon, but topographical reasons are difficult to imagine, and invoking "chance" is equally difficult. On the contrary, either by chance or by design, a very clear astronomical solution does exist.

---
[1] Declinations in this paper are calculated using the program get-Dec kindly provided by C. Ruggles
[2] Paris' paper has been written much before the birth of modern Archaeoastronomy and is pioneristic under many respect, although some of his assertions and calculations must be taken with outmost care. The present author is currently carrying out a coomprenshive re-examination of the Archaeoastronomy of all the Angkor complexes.

As is well known, the plane containing the Earth and the Moon orbit is not the ecliptic, but forms with it an angle of 5° 9'. This has the consequence that the maximal and minimal declinations which the Moon can attain are greater/lesser than those of the sun (which of course equal ± the obliquity of the ecliptic) by such an amount. This leads to the fact that the Moon at the horizon can attain azimuths lesser/greater than those of the sun at the solstices; due to a series of physical reasons however the extreme declinations, also called maximal standstills, are attained only once each every 18.6 years. Of particular interest is the full Moon closest in time to the winter solstice, since it always attains a declination close to the maximal one in the year of the standstill. The full Moon near the winter solstice – the longest night of the year – thus culminates very high in the sky (higher than the Sun at summer solstice, in particular) and remains in the sky the longest.

Preah Khan of Kompong Svay is definitively oriented to the Moon rising at the maximal northern standstill. Indeed azimuth 60° with a flat horizon at this latitude yields a declination +28° 56' (the true lunar declination at the site is actually slightly greater, 29° 11', because parallax must be taken into account). This value has to be compared with the standstill declination of +28° 39': again, as is the case of Koh Ker for the zenith passages, the matching is really impressive.[3] Of course the role of the Moon is quite relevant both in Hinduism – where it is identified with the God Chandra – and in Buddhism, since festivals and recurrences associated with Buddha's life are timed by the full Moon.

There is a second point perhaps related to astronomy which is worth discussing about Preah Khan of Kompong Svay. As already known in the literature, the temple is located on the same latitude of Angkor Vat. The accuracy of this coincidence is astonishing, the error being less than one arcminute. The following is indeed true (data obtained with Google Earth Pro): the center of the Angkor Vat temple is within 1' at the same latitude (13° 24' ) of the entrance to the Preah Khan inner enclosure. In between, the same parallel passes 150 meters to the south of the south boundary of the Chau Srei Vibol temple, which is located 17 Kms east of Angkor and was probably constructed in the 11[th] century as well.

It is, again, possible to think that all this is due to a chance. However, there exist a impressive series of linear patterns which connect the sides and the centers of many of the temples in Angkor (Paris 1941) so it is not inconceivable that this is rather a symbolic link designed by the architects of one, or even of both, monuments. If it is so, how did they manage to realize it?

The earth is round, so that "tracing lines" on a map to connect ancient monuments can be a very dangerous procedure, as inter-visibility must be taken into account (see e.g. Magli 2013, 2015). The visibility of an object which is h meters high equals the square root of 13 h expressed in kilometers (so that, for instance, a person 2 meters tall sees on a flat horizon at about 5 Kms distance). This means that the Angkor Vat central tower – which reaches 65 meters above ground – was visible from Chau Srei Vibol, which is located on the same flatland; to adjust the surveyor's east-west line the architect had to choose an approximate point, to find the meridian at this point (using the Sun and the well known "Indian circle" method, or the stars) and then adjust his position by repeating the measures until the tower at the horizon was collimated at true east: by far *not* an easy task, but possible.

Quite another story is to connect two non-intervisible points, a thing which generally has no sense at all since of course there are no straight lines on a curved surface. However, the special case of sites placed at the same latitude makes sense, because – although the parallel circle is *not* the shortest path between two points at the same latitude - the parallel in itself is (in principle) easy to determine using astronomy. It is indeed possible to establish latitude by measuring the height of the celestial pole (which however was in a dark zone in that period, due to precession, so only an indirect method based on the rotation of stars could be applied) or the height of the sun at midday in fixed days. This second method would seem favored, also because of the orientation of Koh Ker which suggests accurate observations of the zenith sun. In any case, obtaining voluntarily the accuracy exhibited by the Angkor- Preah Khan alignment must have been quite a daunting task.

---

[3]The variation of the obliquity of the ecliptic since 1100 AD is negligible, being about 6'.

**4. Discussion**

In the above I have discussed three true facts: the Koh Ker pyramid is oriented to the rising sun on the days of the zenith passages, the Preah Khan complex is oriented to the Moon rising at the maximal northern standstill and finally, the latter complex lies on the same parallel of Angkor Vat. I did not, however, show that these facts are not due to a chance. Of course indeed, the mere existence of astronomical alignments does not show that they are deliberate, and a proof might be obtained only trough a accurate analysis of the Khmer period – including contemporary texts and inscriptions - which is far beyond the expertise of the present author.
My hope is, however, that the present archaeo-astronomical study can contribute to the interpretation of these magnificent monuments.